# Markov chain random fields, spatial Bayesian networks, and optimal neighborhoods for simulation of categorical fields

Weidong Li, Chuanrong Zhang

*Department of Geography & Center for Environmental Science and Engineering, University of Connecticut, Storrs, CT 06269, USA.*

Correspondence: Weidong Li (email: weidong.li@uconn.edu)

**Abstract**
The Markov chain random field (MCRF) model/theory provides a non-linear spatial Bayesian updating solution at the neighborhood nearest data level for simulating categorical spatial variables. In the MCRF solution, the spatial dependencies among nearest data and the central random variable is a probabilistic directed acyclic graph that conforms to a neighborhood-based Bayesian network on spatial data. By selecting different neighborhood sizes and structures, applying the spatial conditional independence assumption to nearest neighbors, or incorporating ancillary information, one may construct specific MCRF models based on the MCRF general solution for various application purposes. Simplified MCRF models based on assuming the spatial conditional independence of nearest data involve only spatial transition probabilities, and one can implement them easily in sequential simulations. In this article, we prove the spatial Bayesian network characteristic of MCRFs, and test the optimal neighborhoods under the spatial conditional independence assumption. The testing results indicate that the quadrantal (i.e., one nearest datum per quadrant) neighborhood is generally the best choice for the simplified MCRF solution, performing better than other sectored neighborhoods and non-sectored neighborhoods with regard to simulation accuracy and pattern rationality.

## 1. Introduction

Simulation of heterogeneous categorical spatial variables (or categorical fields) is crucial for many purposes, such as spatial or spatiotemporal mapping and uncertainty assessment of various surface and subsurface landscapes in complex geospatial systems [16, 15, 38]. The Markov chain random field (MCRF) model/theory, which was proposed in recent years [17, 20, 22], provides a non-linear fundamental spatial Bayesian updating solution at the neighborhood nearest data level for simulating categorical spatial variables. Because the MCRF model is a generalized fundamental spatial model, it can serve as the base model of more integrative models for various application purposes. Therefore, a nonlinear geostatistical approach (i.e., the MCRF approach or Markov chain geostatistics) was proposed based on the MCRF model/theory for simulating categorical spatial variables. The general full solution of MCRFs represents a Bayesian updating-based multiple-point spatial statistical model. Based on the *spatial conditional independence* assumption for nearest spatial data within a neighborhood, the general full solution of MCRFs can be largely simplified to a form that comprises only spatial transition probabilities, which can be estimated from sparse sample data through transiogram modeling [18]. A random-path sequential class simulation algorithm for the simplified MCRF solution demonstrated obvious advantages over the conventional kriging-based sequential indicator simulation algorithm [19]. A typical characteristic of the MCRF approach is that it can generate aggregative spatial patterns, which is not only preferred for categorical mapping but also can reproduce input spatial transition probabilities. An extension has been developed for incorporating ancillary data through cosimulation [22], and the MCRF cosimulation (coMCRF) model has proved to be effective in post-processing land cover/use data pre-classified from remote sensed imagery for land cover/use study and urban growth detection [39-41]. Therefore, the MCRF approach holds great promise as a versatile spatial statistical approach for dealing with complex categorical spatial variables.

A MCRF is considered as a spatial Markov chain that moves or jumps in a space, locally conditioned through sequential Bayesian updating over nearest data within local neighborhoods. Its local conditional probability distribution can be well-explained by a local sequential Bayesian updating process within a neighborhood [22], thus applying Bayesian updating [7, 14] to individual spatial data within a neighborhood. Therefore, the MCRF model is not a conventional Markov random field (MRF) model as described by Besag [3] in the MRF theory or used by Norberg *et al*. [27] in geology, although it represents an extension of the conventional Markov chain theory toward a multi-dimensional space. Rather, the MCRF model represents a probabilistic directed acyclic graphical model over a neighborhood of spatial data and the central random variable, or simply a neighborhood-based *spatial Bayesian network*, conforming to the conventional Bayesian networks [29] that have been widely used for non-spatial data, such as in medicine or ecosystem service [12].

The MCRF sequential class simulation algorithm has been using a quadrantal neighborhood, which considers only one nearest neighbor from each quadrant that sectors the nearest-data search area. The rationale behind this neighborhood is that it may approximately meet (or have least deviation from) the requirement for conditional independence of nearest data in a neighborhood [17, 19]. Such a spatial conditional independence assumption for nearest data is supported by the cardinal-neighbor conditional independence property of Pickard random fields [31, 33] – given the state of the central cell adjacent neighbors in cardinal directions on a lattice are conditionally independent – which was extended to the sparse data situation [17]. In addition, neighborhood sectorization (mainly quadrant or octant search neighborhoods) has been a strategy used in geostatistics [13] or in other places for reducing the data clustering effect. However, the MCRF model does not limit its neighborhood choice to the quadrantal form. Neighborhoods without sectorization or with other sector numbers may be practically feasible under the spatial conditional independence assumption, although they may deviate more from the conditional independence of nearest neighbors compared to the quadrantal neighborhood structure. Thus, a test on other neighborhood choices may provide insights into the effect of neighborhood deviation from conditional independence on simulation accuracy.

In this article, we aim to: (i) prove that the MCRF model (both full and simplified solutions) is essentially a spatial Bayesian network over a neighborhood of spatial data and the central unobserved location; and (ii) explore the effect of neighborhood size and sectorization on simulation accuracy of the simplified MCRF model under the spatial conditional independence assumption. We performed a series of test simulations to see whether there are some optimal neighborhood sizes and structures with the simplified MCRF model. Such a study may improve the understanding about the MCRF model and guide the development and application of simulation algorithms based on the simplified MCRF model.

## 2. Markov chain random fields

### 2.1. General full solution

The basic idea for the MCRF model is that a spatial Markov chain moving or jumping in a space is locally conditioned on nearest data in different directions within local neighborhoods through sequential Bayesian updating over individual nearest data serving as new evidence (Fig. 1) [17, 20, 22].

The MCRF model was proposed mainly for simulating categorical fields conditioned on sample data, even though it may work with lattice data and might be applicable to continuous spatial variables. Nonetheless, here we only consider categorical fields and sample data. Assume that $Z\{Z(\mathbf{u}): \mathbf{u} \in D\}$ is a finite categorical random field on a fixed subset $D$ of a $d$-dimensional space $R^d$ with a state space $S = (1, \ldots n)$, in which $M$ sampled locations are observed and the left $N$ unobserved locations are subject to estimation. Assume a Markov-type neighborhood (i.e., only consider nearest data nodes in different directions around an unobserved location). Let $\{i_k(\mathbf{u}_k), k = 1, \ldots, m\}$ represent the $m$ nearest data in different directions within a neighborhood around an unobserved location $\mathbf{u}_0$, and $i_0(\mathbf{u}_0)$ ($i_0 = 1, \ldots, n$) represents the possible value (i.e., state or class label) of the random variable $Z(\mathbf{u}_0)$ to be estimated at the unobserved location $\mathbf{u}_0$. Then, the univariate conditional probability of $Z\{Z(\mathbf{u}): \mathbf{u} \in D\}$ at the unobserved location $\mathbf{u}_0$ is localized as

$$f_Z[Z(\mathbf{u}_0) = i_0(\mathbf{u}_0)|Z(D_{/\mathbf{u}_0})] = p[i_0(\mathbf{u}_0)|i_1(\mathbf{u}_1), \ldots, i_m(\mathbf{u}_m)], \tag{1}$$

where $Z(D_{/\mathbf{u}_0})$ denotes the entire finite categorical random field (including observed and unobserved nodes) excluding the location $\mathbf{u}_0$. Thus, the categorical random field is actually defined on unobserved locations, locally conditioned on nearest data with an unfixed neighborhood system (i.e., the number of nearest data and their locations are unfixed). Then we have the joint probability distribution of the whole categorical random field $Z\{Z(\mathbf{u}): \mathbf{u} \in D\}$ to be

$$F_Z[(N), (M)] = F_Z[(N)|(M)]F_Z[(M)], \tag{2}$$

where $(N)$ denotes the $N$ random variables at the $N$ unobserved locations, $(M)$ denotes the $M$ sample data at the $M$ observed locations, and $F_Z[(M)]$ is a constant. The joint probability distribution of the $N$ random variables, conditioned on the $M$ sample data, is the product of the specific local conditional probabilities at the $N$ unobserved locations:

$$F_Z[(N)|(M)] = \prod_{k=1}^{N} p_k[i_{0_k}(\mathbf{u}_{0_k})|i_{1_k}(\mathbf{u}_{1_k}), \ldots, i_{m_k}(\mathbf{u}_{m_k})]. \tag{3}$$

To conduct conditional simulations, however, what we need to know is how to solve the generalized local conditional probability distribution - $p[i_0(\mathbf{u}_0)|i_1(\mathbf{u}_1), \ldots, i_m(\mathbf{u}_m)]$.

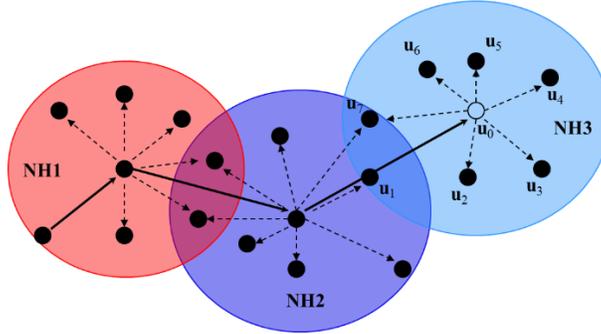

**Fig. 1**. Illustration of a MCRF with different neighborhoods at different unobserved locations (based on the figure 1 in Li [17], in which dependencies between nearest data were not shown).

According to the relationship of conditional probability and joint probability, we have

$$p[i_0(\mathbf{u}_0)|i_1(\mathbf{u}_1), \ldots, i_m(\mathbf{u}_m)] = \frac{p[i_0(\mathbf{u}_0), i_1(\mathbf{u}_1), \ldots, i_m(\mathbf{u}_m)]}{p[i_1(\mathbf{u}_1), \ldots, i_m(\mathbf{u}_m)]}. \tag{4}$$

Emphasizing the single chain nature of a MCRF and the last visited location, we can factorize the joint probability $p[i_0(\mathbf{u}_0), i_1(\mathbf{u}_1), \ldots, i_m(\mathbf{u}_m)]$ in the right hand side of equation (4), that is, the joint probability of the neighborhood $\{i_1(\mathbf{u}_1), \ldots, i_m(\mathbf{u}_m)\}$ and the central random variable, as

$$p[i_0(\mathbf{u}_0), i_1(\mathbf{u}_1), \ldots, i_m(\mathbf{u}_m)] =$$
$$p[i_m(\mathbf{u}_m)|i_0(\mathbf{u}_0), \ldots, i_{m-1}(\mathbf{u}_{m-1})] \cdots p[i_2(\mathbf{u}_2)|i_0(\mathbf{u}_0), i_1(\mathbf{u}_1)] \, p[i_0(\mathbf{u}_0)|i_1(\mathbf{u}_1)] \, p[i_1(\mathbf{u}_1)], \quad (5)$$

where $p[i_1(\mathbf{u}_1)]$ is a constant value. Thus, this factorization means that the joint probability is actually a result of an updating process of a transition probability $p[i_0(\mathbf{u}_0)|i_1(\mathbf{u}_1)]$ over nearest data $i_2(\mathbf{u}_2)$, …, $i_m(\mathbf{u}_m)$ within a neighborhood. Combining equations (4) and (5), we have the generalized local conditional probability distribution decomposed as

$$p[i_0(\mathbf{u}_0)|i_1(\mathbf{u}_1), \ldots, i_m(\mathbf{u}_m)] =$$
$$\frac{1}{A_1} p[i_m(\mathbf{u}_m)|i_0(\mathbf{u}_0), \ldots, i_{m-1}(\mathbf{u}_{m-1})] \cdots p[i_2(\mathbf{u}_2)|i_0(\mathbf{u}_0), i_1(\mathbf{u}_1)] \, p[i_0(\mathbf{u}_0)|i_1(\mathbf{u}_1)], \quad (6)$$

where $A_1 = p[i_1(\mathbf{u}_1), \ldots, i_m(\mathbf{u}_m)] / p[i_1(\mathbf{u}_1)]$ is a constant because it does not involve the unobserved location $\mathbf{u}_0$, and $\mathbf{u}_1$ is the last visited location or the location that the spatial Markov chain goes through to the current unobserved location $\mathbf{u}_0$ [17]. This factorization process is in accordance with Bayes' theorem [2]. Equation (6) characterizes the neighborhood situations NH1 and NH3 in Fig. 1.

There are two ways to factorize $p[i_1(\mathbf{u}_1), i_0(\mathbf{u}_0)]$ in the factorization process of the joint probability $p[i_0(\mathbf{u}_0), i_1(\mathbf{u}_1), \ldots, i_m(\mathbf{u}_m)]$ - It can be factorized either as $p[i_0(\mathbf{u}_0)|i_1(\mathbf{u}_1)] \, p[i_1(\mathbf{u}_1)]$ or as $p[i_1(\mathbf{u}_1)|i_0(\mathbf{u}_0)] \, p[i_0(\mathbf{u}_0)]$. Consequently, we can rewrite equation (5) as

$$p[i_0(\mathbf{u}_0), i_1(\mathbf{u}_1), \ldots, i_m(\mathbf{u}_m)] =$$
$$p[i_m(\mathbf{u}_m)|i_0(\mathbf{u}_0), \ldots, i_{m-1}(\mathbf{u}_{m-1})] \cdots p[i_2(\mathbf{u}_2)|i_0(\mathbf{u}_0), i_1(\mathbf{u}_1)] \, p[i_1(\mathbf{u}_1)|i_0(\mathbf{u}_0)] \, p[i_0(\mathbf{u}_0)], \quad (7)$$

which is more commonly used. Equation (7) means that the joint probability $p[i_0(\mathbf{u}_0), i_1(\mathbf{u}_1), \ldots, i_m(\mathbf{u}_m)]$ is a result of an updating process of a marginal probability $p[i_0(\mathbf{u}_0)]$ over nearest data $i_1(\mathbf{u}_1)$, …, $i_m(\mathbf{u}_m)$ within a neighborhood. Combining equations (4) and (7), then we have the generalized local conditional probability distribution decomposed as

$$p[i_0(\mathbf{u}_0)|i_1(\mathbf{u}_1), \ldots, i_m(\mathbf{u}_m)] =$$
$$\frac{1}{A_2} p[i_m(\mathbf{u}_m)|i_0(\mathbf{u}_0), \ldots, i_{m-1}(\mathbf{u}_{m-1})] \cdots p[i_2(\mathbf{u}_2)|i_0(\mathbf{u}_0), i_1(\mathbf{u}_1)] \, p[i_1(\mathbf{u}_1)|i_0(\mathbf{u}_0)] \, p[i_0(\mathbf{u}_0)], \quad (8)$$

where $A_2 = p[i_1(\mathbf{u}_1), \ldots, i_m(\mathbf{u}_m)]$ is a constant, and $\mathbf{u}_1$ is just a nearest neighbor, serving as the start datum for sequential Bayesian updating. Equation (8) characterizes the neighborhood situation NH2 in Fig. 1.

Equation (6) is regarded as the general full solution of the local conditional probability distributions of MCRFs. From the Bayesian perspective, $p[i_0(\mathbf{u}_0)|i_1(\mathbf{u}_1), \ldots, i_m(\mathbf{u}_m)]$ is the posterior probability, $p[i_0(\mathbf{u}_0)|i_1(\mathbf{u}_1)]$ is the prior probability (i.e., a Markov chain transition probability), and the remaining part of the right hand side of equation (6) excluding the constant is the likelihood component. The likelihood component comprises multiple three-point to $m+1$-point likelihoods, each for one nearest datum except for the last visited location. These likelihoods for spatial data, from $p[i_m(\mathbf{u}_m)|i_0(\mathbf{u}_0), \ldots, i_{m-1}(\mathbf{u}_{m-1})]$ to $p[i_2(\mathbf{u}_2)|i_0(\mathbf{u}_0), i_1(\mathbf{u}_1)]$, represent a process of sequential Bayesian updating based on different nearest data within a neighborhood. Each nearest datum represents new evidence that may be used to update the previous information about the local conditional probability at location $\mathbf{u}_0$ to make it closer to the truth. The updating process starts from nearest neighbor $i_2(\mathbf{u}_2)$ and ends at nearest neighbor $i_m(\mathbf{u}_m)$ within the neighborhood around the unobserved location $\mathbf{u}_0$. Nearest neighbors used in earlier updatings in the updating sequence become the conditioning data of later updatings, and all updatings are conditioned on $i_0(\mathbf{u}_0)$ being estimated, as illustrated in Fig. 2(a), with a neighborhood of six nearest neighbors. Such a sequential Bayesian updating on nearest spatial data within a neighborhood is one of the major characteristics of MCRFs. Please note that such a sequential Bayesian updating process is just an interpretation to the model, rather than an iterative computation process in simulation.

Compared with equation (6), equation (8) has a similar sequential Bayesian updating process, but its prior probability becomes a marginal probability $p[i_0(\mathbf{u}_0)]$, and there is one more likelihood - $p[i_1(\mathbf{u}_1)|i_0(\mathbf{u}_0)]$, as illustrated in Fig. 2(b) with a neighborhood of six nearest neighbors. Based on the detailed balance principle, we have $p[i_0(\mathbf{u}_0)|i_1(\mathbf{u}_1)]\,p[i_1(\mathbf{u}_1)] = p[i_1(\mathbf{u}_1)|i_0(\mathbf{u}_0)]\,p[i_0(\mathbf{u}_0)]$. Thus, given the same number of nearest neighbors, equation (8) is equivalent to equation (6). If we assume that the spatial Markov chain always goes through a nearest neighbor to reach the current location $\mathbf{u}_0$ being estimated, equation (6) is sufficient to be the general full solution of MCRFs. Another reason for choosing equation (6) as the general full solution of MCRFs is that it more clearly embodies the locally-conditioned single-Markov-chain characteristic of MCRFs. In addition, equation (8) also can be explained in this way: The last visited location is far away from the current location (i.e., outside the neighborhood), and its impact may be ignored due to the long distance and the screening effect of nearest data, or the transition probability over a long separate distance decays to the marginal probability $p[i_0(\mathbf{u}_0)]$.

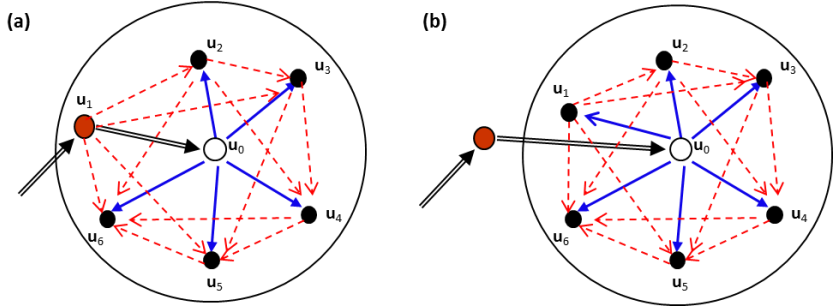

**Fig. 2.** Data dependencies in local neighborhoods of MCRFs: (**a**) A MCRF neighborhood with six nearest neighbors and data dependencies (assuming $\mathbf{u}_1$ is the last visited location of the spatial Markov chain). (**b**) The same neighborhood, but the last visited location is assumed to be far away (not in the neighborhood). The data dependencies also indicate the sequential Bayesian updating process. Data dependencies across the unobserved location $\mathbf{u}_0$ being estimated are ignored based on Markov property and for clarity of illustration.

*2.2. General simplified solution based on spatial conditional independence assumption*

In equations (6) and (8), the likelihood components consist of a series of three-point to $m+1$-point likelihoods that involve the unobserved location $\mathbf{u}_0$. Thus, the general full solution of MCRFs actually represents a spatial model of multiple-point statistics (here multiple-point likelihoods) at the neighborhood nearest data level. These multiple-point statistics are functions of separate distances because these data points may not be immediately adjacent in a sparse or inexhaustive data space. Without simplification, the multiple-point likelihoods are difficult to estimate from sparse sample data.

Conditional independence is a basic property of one-dimensional Markov chains and it is a practical assumption widely used in Bayesian analysis of non-spatial data [8, 30]. For real spatial data, spatial conditional independence of nearest neighbors within a neighborhood rarely holds. Even so, practical use of the conditional independence assumption to spatial data is still feasible for simplifying models or computation, if the spatial structures of neighborhoods are not strongly asymmetric, as suggested in Li [17] and used in Li and Zhang [19]. In fact, this assumption theoretically holds in Pickard random fields for adjacent cells in cardinal directions given the value of the central cell [31, 33]. The *spatial* conditional independence assumption used here is an extension of the cardinal-neighbor conditional independence property of Pickard random fields toward the sparse spatial data situation [17], and it assumes that given the value of the random variable at the central location $\mathbf{u}_0$ its nearest neighbors in different directions are conditionally independent. Thus, for the neighborhood $\{i_1(\mathbf{u}_1), \ldots, i_m(\mathbf{u}_m)\}$ of the unobserved location $\mathbf{u}_0$ in the categorical random field $Z\{Z(\mathbf{u}): \mathbf{u} \in D\}$, we have

$$p[i_2(\mathbf{u}_2)|i_0(\mathbf{u}_0), i_1(\mathbf{u}_1)] = p[i_2(\mathbf{u}_2)|i_0(\mathbf{u}_0)]$$
$$\ldots$$

$$p[i_m(\mathbf{u}_m)|i_0(\mathbf{u}_0), \ldots, i_{m-1}(\mathbf{u}_{m-1})] = p[i_m(\mathbf{u}_m)|i_0(\mathbf{u}_0)]. \tag{9}$$

With such a conditional independence assumption for spatial data within a neighborhood, all of the multi-point likelihoods in equation (6) become two-point likelihoods, and equation (6) is simplified to

$$p[i_0(\mathbf{u}_0)|i_1(\mathbf{u}_1), \ldots, i_m(\mathbf{u}_m)] = \frac{1}{A_1} p[i_m(\mathbf{u}_m)|i_0(\mathbf{u}_0)] \cdots p[i_2(\mathbf{u}_2)|i_0(\mathbf{u}_0)] p[i_0(\mathbf{u}_0)|i_1(\mathbf{u}_1)] \tag{10}$$

Fig. 3(a) illustrates this simplified MCRF solution with a neighborhood of six nearest neighbors. Similarly, equation (8) is simplified to

$$p[i_0(\mathbf{u}_0)|i_1(\mathbf{u}_1), \ldots, i_m(\mathbf{u}_m)] = \frac{1}{A_2} p[i_m(\mathbf{u}_m)|i_0(\mathbf{u}_0)] \cdots p[i_1(\mathbf{u}_1)|i_0(\mathbf{u}_0)] p[i_0(\mathbf{u}_0)], \tag{11}$$

which is illustrated by Fig. 3(b).

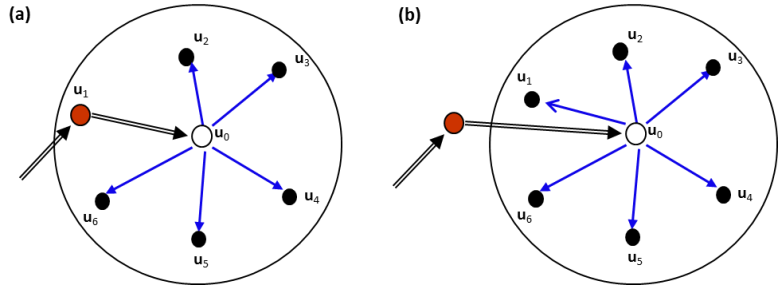

**Fig. 3.** Data dependencies in local neighborhoods of simplified MCRFs under the spatial conditional independence assumption of nearest data: (**a**) A MCRF neighborhood with six nearest neighbors and data dependencies (assuming $\mathbf{u}_1$ is the last visited location of the spatial Markov chain). (**b**) The same neighborhood, but the last visited location is assumed to be far away (not in the neighborhood). The data dependencies only show the transition probability directions.

Due to the complex correlations among spatial data, equations (10) and (11) do not truly hold for real data. The extents of their departures from the real conditional independence largely depend on the numbers and spatial configurations of nearest data within local neighborhoods. When the neighborhood size is large and nearest data are clustered, the above spatial conditional independence assumption should largely deviate from the reality; but when the neighborhood size is small and nearest data are non-clustered, the above spatial conditional independence assumption should be close to the reality. Ripley [32] suggested that in a pairwise interaction process site interactions between a point and its distant nearest neighbors (in any directions) might be treated independently. As we tested in this article, although the conditional independence assumption can work with spatial data, there are optimal choices in neighborhood size and sectorization.

*2.3. General simplified solution with transition probability notation*

The two-point likelihoods and conditional probability in the right hand side of equation (10) and equation (11) are all spatial transition probabilities over their respective separate distances (i.e., spatial lags). For example, $p[i_m(\mathbf{u}_m)|i_0(\mathbf{u}_0)]$ can be written as $p_{i_0 i_m}(\mathbf{h}_{0m})$ with $\mathbf{h}_{0m} = \mathbf{u}_m - \mathbf{u}_0$. For a continuous variable $\mathbf{h}_{0m}$, $p_{i_0 i_m}(\mathbf{h}_{0m})$ becomes a spatial transition probability function. With the assumption of second-order spatial stationarity, such a spatial transition probability function does not depend on specific spatial locations and may be used as a measure for describing the spatial correlation of categorical spatial data, similar to the variogram. For the convenience of description and following the variogram concept in classical geostatistics [26], the transition probability function over a spatial lag variable, generally written as $p_{ij}(\mathbf{h})$, was named as the "*transiogram*" [18]. Thus, a transiogram is theoretically expressed as

$$p_{ij}(\mathbf{h}) = \Pr[Z(\mathbf{u}+\mathbf{h}) = j \mid Z(\mathbf{u}) = i], \qquad (12)$$

where $i$ and $j$ denote the specific states (i.e., classes) of categorical random variables $Z(\mathbf{u})$ and $Z(\mathbf{u}+\mathbf{h})$ at locations $\mathbf{u}$ and $\mathbf{u} + \mathbf{h}$, respectively. Similar to the variogram, the transiogram $p_{ij}(\mathbf{h})$ is graphically a curve with increasing $\mathbf{h}$ from zero. The transiogram expression in equation (12) is identical to the spatial transition probability function in Luo [24]. It is well known that Carle and Fogg [5] applied similar spatial transition probability functions to indicator kriging/cokriging systems for simulating hydrofacies in three dimensions, as implemented in Weissimann et al. [36].

With the spatial second-order stationarity assumption and the transition probability notation, as well as the law of total probability, equation (10) can be rewritten as

$$p[i_0(\mathbf{u}_0)|i_1(\mathbf{u}_1), \ldots, i_m(\mathbf{u}_m)] = \frac{p_{i_1 i_0}(\mathbf{h}_{10}) \prod_{g=2}^{m} p_{i_0 i_g}(\mathbf{h}_{0g})}{\sum_{f_0=1}^{n}[p_{i_1 f_0}(\mathbf{h}_{10}) \prod_{g=2}^{m} p_{f_0 i_g}(\mathbf{h}_{0g})]} \qquad (13)$$

where $p_{i_0 i_g}(\mathbf{h}_{0g})$ represents a specific transition probability from class $i_0$ to class $i_g$ over the spatial lag $\mathbf{h}_{0g}$ between locations $\mathbf{u}_0$ and $\mathbf{u}_g$, which can be drawn from a transiogram model $p_{i_0 i_g}(\mathbf{h})$; $i_1(\mathbf{u}_1)$ represents the nearest data neighbor from or across which the spatial Markov chain moves to the current unobserved location $\mathbf{u}_0$; $m$ represents the number of nearest data including the last visited location within a neighborhood; and $i$ and $f$ represent states (i.e., classes) in the state space $S = (1, \ldots, n)$ of the MCRF. Equation (13) represents the general simplified solution of MCRFs with the transition probability notation and the second-order stationarity assumption. It is still in accordance with the Bayesian updating formula (6), with $p[i_0(\mathbf{u}_0)|i_1(\mathbf{u}_1), \ldots, i_m(\mathbf{u}_m)]$ being the posterior probability, $p_{i_1 i_0}(\mathbf{h}_{10})$ being the prior probability, the denominator being the normalizing constant, and the remaining part of the right hand side being the likelihood component.

Similarly, equation (11) can be rewritten using the transition probability notation with the spatial second-order stationarity assumption as

$$p[i_0(\mathbf{u}_0)|i_1(\mathbf{u}_1), \ldots, i_m(\mathbf{u}_m)] = \frac{p_{i_0} \prod_{g=1}^{m} p_{i_0 i_g}(\mathbf{h}_{0g})}{\sum_{f_0=1}^{n}[p_{f_0} \prod_{g=1}^{m} p_{f_0 i_g}(\mathbf{h}_{0g})]} \qquad (14)$$

where $p_{i_0}$ is the marginal probability of class $i_0$. Here $p_{i_0}$ is also assumed to be spatially stationary, and it is actually the class proportion of class $i_0$. In addition, Equation (14) can be directly derived from Equation (13) by using the detailed balance principle, $p_{i_0 i_1}(\mathbf{h}_{01}) p_{i_0} = p_{i_1 i_0}(\mathbf{h}_{10}) p_{i_1}$, which assumes a reversible stationary Markov chain [10]. Equation (14) is still in the Bayesian updating formula, but its prior probability becomes the marginal probability $p_{i_0}$.

Varied forms of MCRF models, especially more complex and integrative models, may be developed based on the full and simplified solutions of MCRFs for various application purposes. For example, the full solution of MCRFs can be partially simplified, coMCRF models can be constructed for incorporating ancillary variables [22], and other information or constraints also can be incorporated for a specific application purpose such as land cover post-classification from remote sensing imagery [40]. In addition, transition probabilities in simplified MCRF models might be replaced by other two-point spatial measures (e.g., indicator covariance) based on their quantitative relationships [5, 25]. However, the transiogram is apparently more intuitive to explain and more flexible in joint model-fitting for multi-class spatial variables than other two-point spatial measures are. By following the variogram concept system, the transiogram concept system provides a large convenience for description and much more clarity for implementing simplified MCRF models.

## 2.4. Relation with Bayesian networks

Although the MCRF model was initially proposed without the knowledge of Bayesian networks [29], we later found that the MCRF model can be regarded as a special spatial Bayesian network over a neighborhood of nearest spatial data and the random variable at the central unobserved location (or a neighborhood-based spatial Bayesian network). Below we address the relationship of MCRFs with Bayesian networks.

Fig. 2(a) and 2(b) were drawn based on equations (6) and (8), respectively, with different prior probability choices. One can see that the general full solution of MCRFs for a local neighborhood can be visualized as a fully-related *probabilistic directed acyclic graph*, similar to a fully-related Bayesian network built on a set of spatial nodes (or locations). After the spatial conditional independence assumption is applied to the nearest data, the general full solution of MCRFs can be simplified into a formula that comprises only spatial transition probabilities. The general simplified solution of MCRFs can be visualized as a pairwise-related *probabilistic directed acyclic graph*, still similar to a pairwise-related Bayesian network built on a set of spatial nodes, as shown in Fig. 3(a) and 3(b) with different prior probability choices. Below we prove that MCRFs conform to neighborhood-based Bayesian networks on spatial data, which may be simply called "*spatial Bayesian networks*" (or probably "*Bayesian random fields*"). Assume that we already had the probabilistic directed acyclic graphs displayed in Figs. 2 and 3 (here redisplayed as spatial Bayesian networks in Fig. 4 for convenience). Let's directly derive the different forms of the MCRF model from these graphs (i.e., spatial Bayesian networks) using the Bayesian network formula [29] - The joint probability distribution of the nodes of a Bayesian network is expressed as the product of all the conditional probabilities (for nodes with parent nodes) and marginal probabilities (for nodes without parent nodes) based on the dependences among the nodes.

Let's use $S_{6+1}$ represent the neighborhood node set of the six nearest data nodes ($\mathbf{u}_1, \mathbf{u}_2, \ldots, \mathbf{u}_6$) and the central random variable node $\mathbf{u}_0$. Based on the spatial dependency relationships of the nodes in the graph of Fig. 4(a), we have the local joint probability distribution for the neighborhood node set $S_{6+1}$ to be directly given as

$$p[i_0(\mathbf{u}_0), i_1(\mathbf{u}_1), \ldots, i_6(\mathbf{u}_6)] = p[i_1(\mathbf{u}_1)] \, p[i_0(\mathbf{u}_0)|i_1(\mathbf{u}_1)] \, p[i_2(\mathbf{u}_2)|i_0(\mathbf{u}_0), i_1(\mathbf{u}_1)] \cdots p[i_6(\mathbf{u}_6)|i_0(\mathbf{u}_0), i_1(\mathbf{u}_1), \ldots, i_5(\mathbf{u}_5)]. \quad (15)$$

Then, we can get the local conditional probability distribution of the random variable $Z(\mathbf{u}_0)$ conditioned on the six nearest data as

$$p[i_0(\mathbf{u}_0)|i_1(\mathbf{u}_1), \ldots, i_6(\mathbf{u}_6)] = \frac{p[i_0(\mathbf{u}_0), i_1(\mathbf{u}_1), \ldots, i_6(\mathbf{u}_6)]}{p[i_1(\mathbf{u}_1), \ldots, i_6(\mathbf{u}_6)]} = \frac{1}{A_3} p[i_6(\mathbf{u}_6)|i_0(\mathbf{u}_0), i_1(\mathbf{u}_1), \ldots, i_5(\mathbf{u}_5)] \cdots p[i_2(\mathbf{u}_2)|i_0(\mathbf{u}_0), i_1(\mathbf{u}_1)] \, p[i_0(\mathbf{u}_0)|i_1(\mathbf{u}_1)], \quad (16)$$

where $A_3 = p[i_1(\mathbf{u}_1), \ldots, i_6(\mathbf{u}_6)] / p[i_1(\mathbf{u}_1)]$ is a constant. If there are $m$ nearest neighbors rather than six, the above equation (16) becomes the general full solution of MCRFs as shown in equation (6). Similarly we can directly get equation (8) from the graph of Fig. 4(b). These mean that the MCRF model is consistent with a Bayesian network over a set of spatial locations considered in a neighborhood before it is simplified using the spatial conditional independence assumption of nearest data within a neighborhood. Consequently, a whole MCRF is globally an extremely complex spatial Bayesian network. Please note that here the six nearest data nodes in the neighborhood node set $S_{6+1}$ are not variable nodes as that used in Pearl [29]. They are just six nodes with known values, and there is only one variable at the central node $\mathbf{u}_0$ that is subjective to estimation.

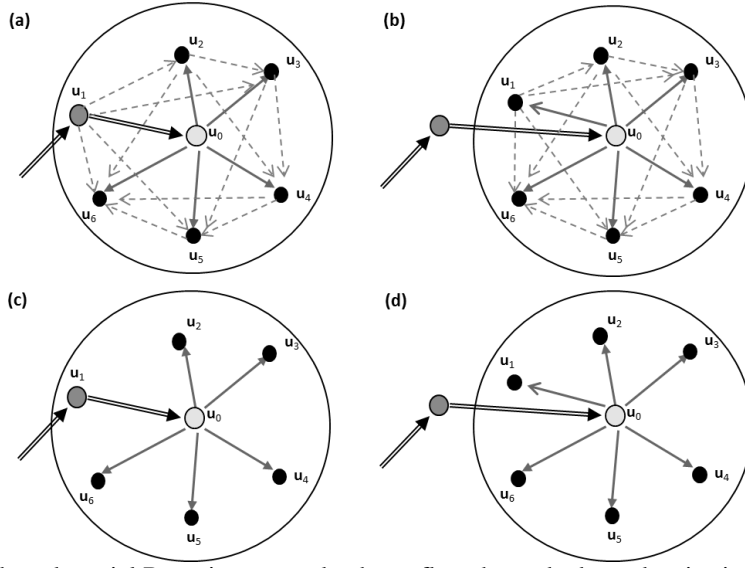

**Fig. 4.** Neighborhood-based spatial Bayesian networks that reflect the node dependencies in local neighborhoods of Markov chain random fields (MCRFs): (**a**) A MCRF neighborhood with six nearest neighbors and node dependencies (assuming $\mathbf{u}_1$ is the last visited location of the spatial Markov chain). (**b**) The same neighborhood, but the last visited location is assumed to be far away (not in the neighborhood). (**c**) A simplified form of (a) based on the spatial conditional independence assumption. (**d**) A simplified form of (b) based on the spatial conditional independence assumption. The node dependencies in (a) and (b) also show the sequential Bayesian updating process. Node dependencies across the uninformed location $\mathbf{u}_0$ being estimated are ignored in (a) and (b) based on Markov property and for clarity of illustration.

Based on the spatial dependency relationships in the graph of Fig. 4(c), we have the local joint probability distribution for the neighborhood node set $S_{6+1}$ to be directly given as

$$p[i_0(\mathbf{u}_0), i_1(\mathbf{u}_1), \ldots, i_6(\mathbf{u}_6)] = p[i_1(\mathbf{u}_1)] \, p[i_0(\mathbf{u}_0)|i_1(\mathbf{u}_1)] \, p[i_2(\mathbf{u}_2)|i_0(\mathbf{u}_0)] \cdots p[i_6(\mathbf{u}_6)|i_0(\mathbf{u}_0)]. \quad (17)$$

Then we can get the local conditional probability distribution of the random variable $Z(\mathbf{u}_0)$ conditioned on the six nearest data as

$$p[i_0(\mathbf{u}_0)|i_1(\mathbf{u}_1), \ldots, i_6(\mathbf{u}_6)] = \frac{p[i_0(\mathbf{u}_0), i_1(\mathbf{u}_1), \ldots, i_6(\mathbf{u}_6)]}{p[i_1(\mathbf{u}_1), \ldots, i_6(\mathbf{u}_6)]} = \frac{1}{A_3} p[i_6(\mathbf{u}_6)|i_0(\mathbf{u}_0)] \cdots p[i_2(\mathbf{u}_2)|i_0(\mathbf{u}_0)] \, p[i_0(\mathbf{u}_0)|i_1(\mathbf{u}_1)], \quad (18)$$

If there are $m$ nearest neighbors rather than six, the above equation (18) becomes the general simplified solution of MCRFs as shown in equation (10). Similarly we can directly get equation (11) from the graph of Fig. 4(d). These mean that the MCRF model is still consistent with a Bayesian network over a set of spatial locations considered in a neighborhood after it is simplified using the spatial conditional independence assumption of nearest data within a neighborhood.

It is obvious that the above MCRF-based spatial Bayesian networks are not the same as traditional Bayesian networks. Traditional Bayesian networks are typically characterized by cause-consequence relationships, where causes are represented by parent nodes and consequences by children nodes [29]. Based on the parent-child relationships in a graph, a probabilistic mathematical model can be obtained. However, there are no clear cause-consequence (or parent-child) relationships among the spatial data nodes and the central unobserved node. This makes it difficult to construct a reasonable and practical Bayesian network on a set of spatial data nodes and unobserved nodes for estimating the random variables on the unobserved nodes. In fact, without equations (6), (8), (10) and (11), we even would not have been able to draw the probabilistic directed acyclic graphs in Figs. 2 and 3 for illustrating the equations in the first place. This is probably why Bayesian networks were traditionally used on non-spatial data. In addition, Bayesian networks do not provide a way to deal with distant interactions of spatial data. Fortunately, the idea of "using a single spatial Markov chain to deal with a whole random field with sample data" plus the idea of "sequential Bayesian updating over nearest data" overcame the challenge of causality [17, 22]. Therefore, in essence, the relationship between the MCRF model and traditional Bayesian networks (including Naive Bayes) is just like the relationship between kriging and multiple linear regression.

It is also obvious that the Bayesian networks of MCRFs in Fig. 4 are not intuitive, rather, they are anti-intuitive to some extent and can only be constructed through sequential Bayesian updating on nearest data within a neighborhood. That is to say, the derivation method of the MCRF model provides a practical way to construct neighborhood-based spatial Bayesian networks on observed and unobserved spatial nodes.

It is worth a mention that using intuition to construct a spatial Bayesian network over a neighborhood node set does not help the estimation of the local conditional probability distribution. The common intuition in spatial statistics or geostatistics is that the nearest neighbors influence the value of the random variable at the central unobserved location. Based on this intuition, the Bayesian network constructed on the same neighborhood node set $S_{6+1}$ should be like that shown in Fig. 5, if we do not consider the dependencies between the six data nodes. In this Bayesian network, the nearest neighbors are parent nodes and the unobserved location is the only child node. Then using the formula of Bayesian networks, we have the joint probability distribution for the neighborhood node set $S_{6+1}$ to be factored as

$$p[i_0(\mathbf{u}_0), i_1(\mathbf{u}_1), \dots, i_6(\mathbf{u}_6)] = p[i_0(\mathbf{u}_0)|i_1(\mathbf{u}_1), \dots, i_6(\mathbf{u}_6)] \prod_{k=1}^{6} p[i_k(\mathbf{u}_k)] . \quad (19)$$

Then, we can get the local conditional probability distribution of the random variable $Z(\mathbf{u}_0)$ conditioned on the six nearest data as

$$p[i_0(\mathbf{u}_0)|i_1(\mathbf{u}_1), \dots, i_6(\mathbf{u}_6)] = \frac{p[i_0(\mathbf{u}_0), i_1(\mathbf{u}_1), \dots, i_6(\mathbf{u}_6)]}{p[i_1(\mathbf{u}_1), \dots, i_6(\mathbf{u}_6)]} = \frac{\prod_{k=1}^{6} p[i_k(\mathbf{u}_k)]}{p[i_1(\mathbf{u}_1), \dots, i_6(\mathbf{u}_6)]} p[i_0(\mathbf{u}_0)|i_1(\mathbf{u}_1), \dots, i_6(\mathbf{u}_6)] . \quad (20)$$

It is apparent that this Bayesian network (Fig. 5(b)) does not help solve any problem; on the contrary, it even creates a self-contradiction, because it requires the nearest data to be completely independent of each other, that is, it requires $p[i_1(\mathbf{u}_1), \dots, i_6(\mathbf{u}_6)] = \prod_{k=1}^{6} p[i_k(\mathbf{u}_k)]$.

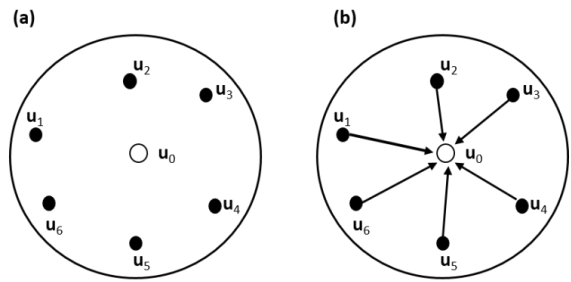

**Fig. 5.** A neighborhood node set (**a**) and a Bayesian network constructed on the neighborhood node set based on the intuition that nearest data influence the value of the random variable at the central unobserved location (**b**). Here $\mathbf{u}_0$ is an unobserved location, and $\mathbf{u}_1$ to $\mathbf{u}_6$ are six nearest data nodes. This Bayesian network based on intuition does not help solve any problem, and it even creates a self-contradiction, as shown in equation (20).

Full independence (i.e., unconditional independence) of nearest data cannot be an assumption in geostatistics or spatial statistics. If the nearest data in a neighborhood $\{i_1(\mathbf{u}_1), \dots, i_6(\mathbf{u}_6)\}$ are unconditionally independent of each other, they should be also independent from the random variable at the central location $\mathbf{u}_0$. Then we have $p[i_0(\mathbf{u}_0), i_1(\mathbf{u}_1), \dots, i_6(\mathbf{u}_6)] = \prod_{k=0}^{6} p[i_k(\mathbf{u}_k)]$ , and $p[i_0(\mathbf{u}_0)|i_1(\mathbf{u}_1), \ldots, i_6(\mathbf{u}_6)] = p[i_0(\mathbf{u}_0)]$, which are not useful estimates based on a local neighborhood of nearest neighbors.

It is also worth a mention that the Bayesian philosophy (e.g., Bayesian updating) has been used in geostatistical modeling previously. However, earlier studies [1, 28, 37] mainly used it for parameter inference or for incorporation of soft information (e.g., expert knowledge, ancillary data), rather than for factorizing a univariate local conditional probability function to specific nearest data as done in Li [17]. By using training images, multiple-point (or high-order) statistics can be estimated from exhaustive data sets for geostatistical mapping of subsurface structures [9, 35]. This may provide a prospect to implement the general full solution of MCRFs. However, using training images may not be practical in many other situations (e.g., complex ground surface landscapes), where real spatial patterns are usually unknown and much more complex but sample data are relatively easy to obtain.

## 3. Testing Different Neighborhoods

### 3.1. Neighborhood choice

The general simplified solution of MCRFs does not consider the effect of data deviation from conditional independence and the effect of data clustering. Apparently, clustered nearest neighbors deviate more from conditional independence than do non-clustered nearest neighbors, and nearest neighbors above the number of cardinal directions deviate more from conditional independence than do nearest neighbors at or below the number of cardinal directions. These situations may affect the estimation accuracy of local conditional probability distributions. Consequently, dealing with these issues in simplified MCRF models may be necessary in algorithm design and neighborhood structure or by adding adjusting parameters. In this paper, we test the effects of different neighborhood sizes and sectorizations with different degrees of deviations from conditional independence on simulation accuracy of a complex categorical landscape pattern using simplified MCRF models.

In real applications, it is unnecessary and difficult to consider many nearest data in different directions for a neighborhood. Those nearest data outside correlation ranges and those approximately located in the same directions with small separate angles but not the closest to the unobserved location $\mathbf{u}_0$ being estimated can be disregarded in equations (13) and (14). Thus, $m$ can be much smaller than the real number of nearest data in different directions. In addition, equations (13) and (14) are based on a generalized spatial conditional independence assumption, which rarely holds for real data, especially for clustered data or neighborhood sizes larger than the number of cardinal directions, because they do not explicitly consider the dependencies and separation angles between nearest data.

An effective method for avoiding data clustering in a neighborhood is sectorization, which divides a search circle or ellipse into a number of sectors (e.g., four or eight) and choose a few nearest data from each sector [13]. The spatial conditional independence assumption can hold for nearest data in cardinal directions by assuming the underlying random field to be a Pickard random field – a stationary Markov field on a finite regular lattice [17]. Therefore, Li and Zhang [19] suggested a random-path MCRF algorithm with a quadrantal neighborhood for sequential class simulations conditioned on point sample data, considering that there are actually four cardinal directions for a discretized field on a square or rectangular lattice. This neighborhood choice relaxes the four cardinal directions to quadrants, by considering four nearest neighbors, one per quadrant, in a search window for estimating the local conditional probability distribution (Fig. 6(d)). The simplified MCRF model for this algorithm is

$$p[i_0(\mathbf{u}_0)|i_1(\mathbf{u}_1), \ldots, i_4(\mathbf{u}_4)] = \frac{p_{i_1 i_0}(\mathbf{h}_{10}) \prod_{g=2}^{4} p_{i_0 i_g}(\mathbf{h}_{0g})}{\sum_{f_0=1}^{n} [p_{i_1 f_0}(\mathbf{h}_{10}) \prod_{g=2}^{4} p_{f_0 i_g}(\mathbf{h}_{0g})]} \tag{21}$$

based on equation (13). In this simplified MCRF model, it is assumed that the last visited location of the spatial Markov chain is always within the four nearest neighbors. Equation (21) assumes spatial conditional independence of the four nearest neighbors in the four quadrants, which approximately represent the four cardinal directions. There may be no data in some quadrants within a search range at boundaries of a study area, or at the beginning of a simulation when sample data are very sparse or a small search radius is used. In such a situation, the size of the neighborhood may occasionally be less than four. Equation (21) can adapt to this situation. In case no data can be found in the entire search circle, it is assumed that the spatial Markov chain comes from a location outside the search range.

In this study, to seek the optimal neighborhood sizes and sector numbers for the simplified solution of MCRFs, and to understand the effect of neighborhood data deviation from conditional independence, we evaluated a series of sectored and non-sectored neighborhoods with different neighborhood sizes and sector numbers. For the non-sectored neighborhoods, nine different neighborhood sizes (i.e., one to nine nearest data) were considered, and the nearest data were obtained within a search radius without considering directions or sectors (Fig. 6(a) and Fig. 6(b)). For sectored neighborhoods, three sector numbers (i.e., two, four and eight) were considered, and only one nearest datum was sought from each sector (Fig. 6(c) to Fig. 6(e)). The directions of sectorization lines may rotate, but that will entail intense computation and may not make much difference in simulated results. Other sector numbers, such as three, five and six, were not considered but their effects on simulated results may be reasoned.

The random-path sequential simulation procedures for all of the non-sectored and sectored neighborhoods are the same, that is, random numbers are used to choose unsampled locations and simulated values during the simulation process and each newly simulated datum is added into the conditioning data set for conditioning later simulations at other unsampled locations. Such a sequential simulation procedure has been widely used in sequential simulation algorithms in geostatistics [11].

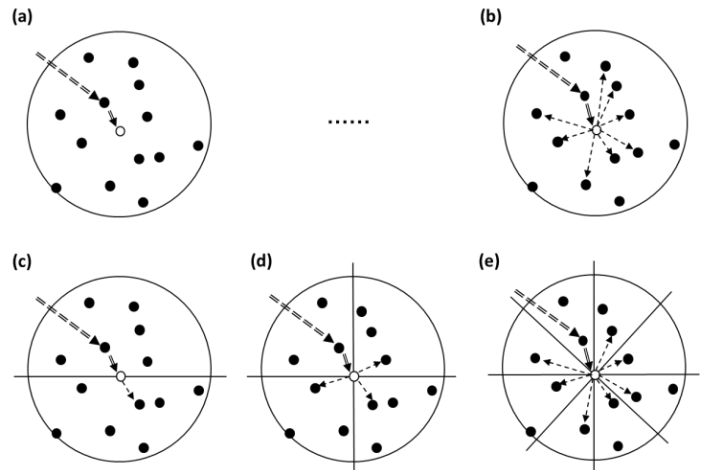

**Fig. 6.** Non-sectored MCRF neighborhoods and sectored MCRF neighborhoods with different neighborhood sizes: (**a**) non-sectored, considering only one nearest neighbor; (**b**) non-sectored, considering nine nearest neighbors; (**c**) two sectors (bisected), considering two nearest neighbors, one per sector; (**d**) four sectors (quadrants), considering four nearest neighbors, one per sector; and (**e**) eight sectors (octants), considering eight nearest neighbors, one per sector. The large circle represents the search area (i.e., moving window) for seeking nearest neighbors used to estimate the unsampled central point (the empty dot). Black solid dots represent locations of data (sample data and previously simulated data) within each neighborhood.

*3.2. Data sets for neighborhood testing*

A complex categorical map with seven classes on a raster field of $128 \times 175$ pixels (total 22400 pixels) was used for testing different MCRF neighborhoods. The map is representative, containing both large classes and minor classes and with apparent non-stationary spatial distribution of different classes. The class names are not our concern in this study, so they are denoted simply as numbers from 1 to 7. This study used four random sample data sets with different densities: sample data set I (646 samples, relatively dense), sample data set II (319 samples, relatively medium), sample data set III (179 samples, relatively sparse), and sample data set IV (89 samples, extra sparse). The reference categorical image and the four sample data sets are provided in Fig. 7. Further sparser sample data sets were not considered because the extra sparse sample data set was already insufficient for transiogram estimation of seven classes.

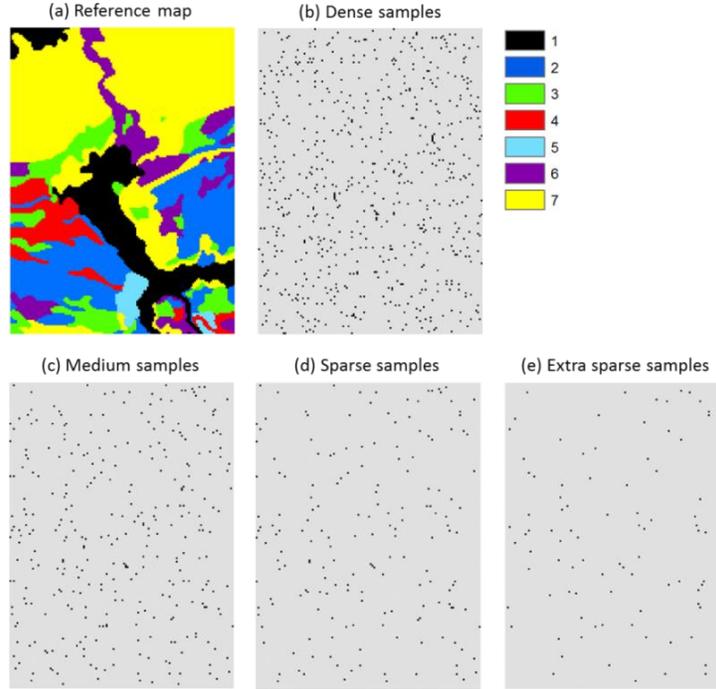

**Fig. 7.** The reference categorical image (**a**) and the four sample data sets (**b** – **e**) used for neighborhood testing

*3.3 Parameter estimation*

To perform simulations using simplified MCRF models, transiogram models are needed to provide transition probability values at the required spatial lags. The estimator for experimental (or empirical) transiograms is given as

$$\hat{p}_{ik}(\mathbf{h}) = \frac{F_{ik}(\mathbf{h})}{\sum_{j=1}^{n} F_{ij}(\mathbf{h})}, \tag{22}$$

where $F_{ik}(\mathbf{h})$ represents the frequency of transitions from class $i$ to class $k$ among sample data pairs with the spatial lag $\mathbf{h}$, and $n$ is the total number of classes in a categorical spatial sample data set. When estimating the transition frequencies from sample data, the lag $\mathbf{h}$ considered is actually a lag interval [$\mathbf{h}$-$\Delta\mathbf{h}$/2, $\mathbf{h}$+$\Delta\mathbf{h}$/2] around $\mathbf{h}$. After experimental transiograms were estimated from sample data, the next step is the joint modeling of experimental transiograms to obtain a complete set of practical transiogram models. Li [17] introduced the transiogram as a generalized two-point spatial correlation measure for categorical data based on the Markov chain theory [10], variogram concepts and models [11], and related pioneer studies [6, 24, 34]. The joint model-fitting issue of experimental transiograms was described in Li and Zhang [21], in which two practical transiogram joint modeling methods were suggested for the MCRF approach. One is a non-parametric method, which linearly interpolates discontinuous experimental transiograms into continuous models. The second is a parametric method, which uses mathematical models to fit experimental transiograms in each row of a transiogram matrix, with one of them to take the left portion of unity. The non-parametric linear interpolation method is highly efficient, and it is suitable when samples are relatively abundant and experimental transiograms are reliable. In addition, Carle and Fogg [6] introduced a transition rate method for transiogram joint modeling. Although the transiogram models produced by this method are relatively oversimplified, it is suitable for use with expert and domain knowledge in subsurface facies simulation, where sample data are usually insufficient for estimating reliable experimental transiograms in horizontal dimensions [18].

Experimental transiograms were estimated from the dense sample data set with a tolerance width of 5 pixel length (e.g., data pairs with lags from 8 to 12 pixel length were counted into the transition frequency for the lag of 10 pixel length) and then interpolated into transiogram models using the linear interpolation method [21]. For comparison, the same set of transiogram models was used for all simulations with different sample data sets so as to exclude the effect of uncertainty in parameter estimation and to focus the testing on the effects of neighborhood structures and data density, although the medium and sparse sample data sets also could provide transiogram models. Two subsets of transiogram models are provided in Fig. 8. The spatial autocorrelation and cross-correlation situations of the related classes can be seen from these transiograms.

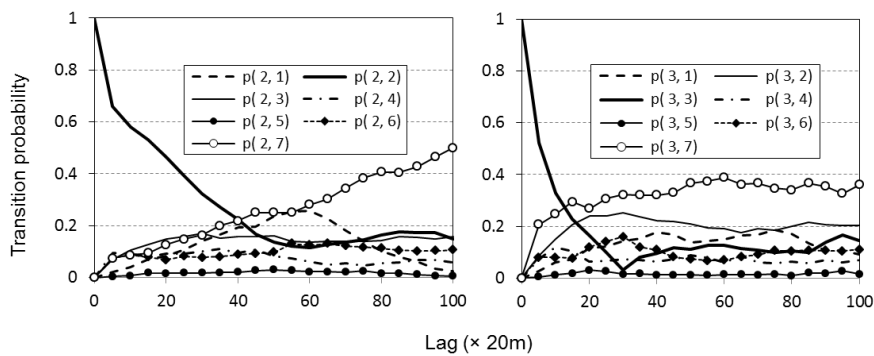

**Fig. 8.** Two subsets of transiogram models obtained by interpolating experimental transiograms estimated from the dense sample data set

*3.4. Simulation outputs*

A total of 48 simulations were performed for the 12 neighborhood choice and 4 sample data sets. For each simulation, 100 realizations were generated, then occurrence probabilities for each class were estimated from the realizations and the optimal prediction map were further obtained based on maximum occurrence probabilities. The search radiuses for these simulations were set to 30 pixel length (for the dense sample data set) to 70 pixel length (for the extra sparse sample data set) to ensure that the required neighborhood sizes can be met at the beginning of simulation except at boundary stripes. Here we use the percentage of correctly simulated locations to represent the accuracy of a simulated realization map or an optimal prediction map (based on maximum occurrence probabilities). Accuracy values were estimated using the reference categorical image as validation data, and accuracy values for simulated realizations were averaged from 100 realizations for each simulation case.

## 4. Testing Results and Discussions

*4.1. Simulation accuracies*

Results (Fig. 9 and Tables 1 and 2) show that the simulation accuracies generally decrease by about 5% for all simulations with the decrease of about 50% in sample density.

Fig. 9(a) and Table 1 provide the simulation accuracy results based on the dense sample data set. For non-sectored neighborhoods, the neighborhood sizes 3, 4 and 5 similarly generate the highest realization accuracy of 75.29%; smaller neighborhood sizes (sizes 1 and 2) generate realizations with much lower average accuracies, and larger neighborhood sizes (sizes 6, 7, 8 and 9) cause the average accuracy of simulated realizations to decrease slightly and gradually with increasing neighborhood sizes. The accuracies of optimal maps are much higher than those of corresponding realizations. Surprisingly, the highest optimal prediction accuracy is obtained by the neighborhood size 2, while the average realization accuracy for this neighborhood size is much lower than those for larger neighborhood sizes. Then the accuracy of optimal prediction gradually decreases with increasing neighborhood size (from size 3 to size 9). Apparently with increasing neighborhood size, the accuracy of optimal prediction based on maximum occurrence probabilities reaches its highest value a little earlier than does the average accuracy of simulated realizations. For sectored neighborhoods, the quadrantal neighborhood has the highest accuracies for both the optimal prediction map and the simulated realizations. The simulation accuracies for the octantal neighborhood are also higher than those for the non-sectored size-8 neighborhood. The bisected neighborhood generates a higher accuracy in optimal prediction than does the non-sectored neighborhood of size 2; however, it does not show any advantage in accuracy of simulated realizations. In general, the simulation accuracy results based on the dense sample data set indicate the following characteristics: (1) while small neighborhood sizes such as 1 and 2 are not good, especially for simulated realizations, too large neighborhood sizes also deteriorate simulation accuracy; (2) sectorization should be a good choice except for the bisected neighborhood; and (3) the quadrantal neighborhood generates the most accurate simulated realizations and optimal prediction map.

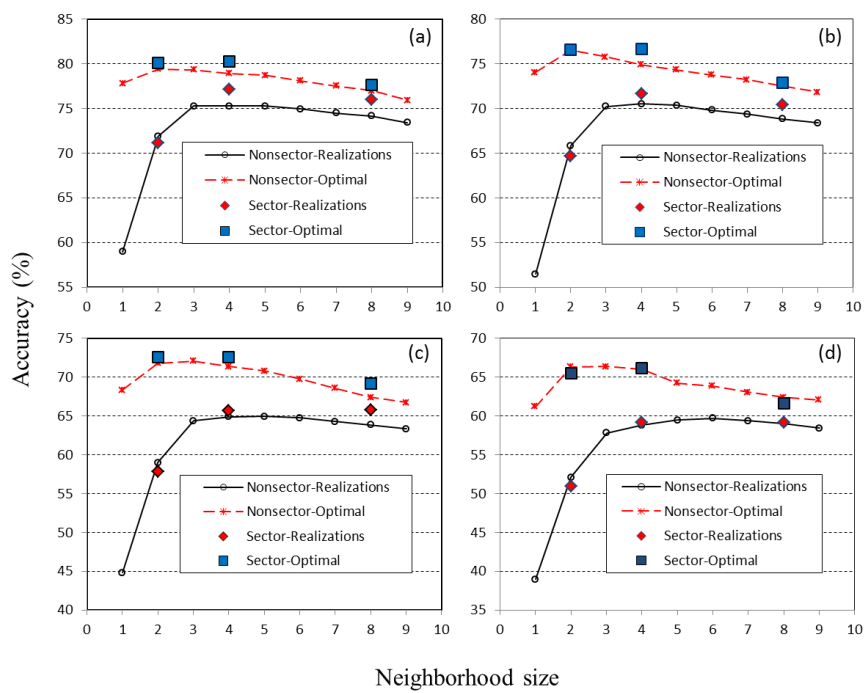

**Fig. 9.** Simulation accuracies of a multi-class categorical map by the MCRF approach using different non-sectored neighborhoods (1 to 9 nearest neighbors) and sectored neighborhoods (2, 4 and 8 sectors, with one nearest neighbor from each sector). (**a**) Dense sample data set (646 data), (**b**) medium sample data set (319 data), (**c**) sparse sample data set (179 data), and (**d**) extra sparse sample data set (89 data). Accuracies for simulated realizations are averaged from 100 realizations for each simulation case. The reference map is used for validation.

**Table 1.** Simulation accuracies with different neighborhood sizes and structures using the dense sample data set (646 random sample data)

|  | Neighborhood size/structure | Realizations* (%) | Optimal** (%) |
|---|---|---|---|
| Non-sectored neighborhood | 1 nearest data | 59.01 | 77.82 |
|  | 2 nearest data | 71.93 | 79.41 |
|  | 3 nearest data | 75.29 | 79.34 |
|  | 4 nearest data | 75.29 | 78.93 |
|  | 5 nearest data | 75.29 | 78.71 |
|  | 6 nearest data | 74.93 | 78.11 |

|  | 7 nearest data | 74.51 | 77.54 |
|---|---|---|---|
|  | 8 nearest data | 74.15 | 77.00 |
|  | 9 nearest data | 73.44 | 75.94 |
| Sectored neighborhood | 2 nearest data /2 sectors | 71.19 | 80.11 |
|  | 4 nearest data /4 sectors | 77.16 | 80.22 |
|  | 8 nearest data /8 sectors | 76.00 | 77.67 |

*Simulation accuracy for realizations was averaged from 100 realizations for each simulation. **Optimal map was based on maximum occurrence probabilities estimated from 100 realizations for each simulation.

The simulation accuracy results based on the medium sample data set are provided in Fig. 9(b). The changes in simulation accuracies with increasing neighborhood sizes and sector numbers for this sample data set have similar trends as for the dense sample data set. There are some minor differences. For example, for the non-sectored neighborhoods, the change in accuracies of optimal prediction maps from neighborhood size 2 to size 9 is a little steeper for this sample data set than that for the dense sample data set; and in this set of simulations the neighborhood size 4 has the highest realization accuracy, which is only a little higher than those with neighborhood sizes 3 and 5. Again, the quadrantal neighborhood performs the best in both simulated realizations and optimal prediction.

For the sparse sample data set, the simulation accuracy results (Fig. 9(c) and Table 2) evince similar trends to those for the dense and medium sample data sets. However, for the non-sectored neighborhoods, the neighborhood size 5 has the highest average accuracy for simulated realizations and neighborhood size 3 has the highest accuracy in optimal prediction, that is, the best neighborhood sizes are a little larger. The quadrantal neighborhood still performs the best; but its advantage over the non-sectored neighborhood of size 4 is less obvious.

**Table 2.** Simulation accuracies with different neighborhood sizes and structures using the sparse sample data set (179 random sample data)

|  | Neighborhood size/structure | Realizations (%) | Optimal (%) |
|---|---|---|---|
| Non-sectored neighborhood | 1 nearest data | 44.82 | 68.36 |
|  | 2 nearest data | 58.97 | 71.79 |
|  | 3 nearest data | 64.35 | 72.08 |
|  | 4 nearest data | 64.87 | 71.38 |
|  | 5 nearest data | 64.95 | 70.79 |
|  | 6 nearest data | 64.74 | 69.76 |
|  | 7 nearest data | 64.27 | 68.58 |
|  | 8 nearest data | 63.87 | 67.38 |
|  | 9 nearest data | 63.31 | 66.75 |
| Sectored neighborhood | 2 nearest data /2 sectors | 57.81 | 72.56 |
|  | 4 nearest data /4 sectors | 65.68 | 72.58 |
|  | 8 nearest data /8 sectors | 65.76 | 69.17 |

The extra-sparse sample data set was used to explore the effect of sample data density on neighborhood sizes and structures when sample data are insufficient. This situation should not occur in real applications because the sample data set is too small relative to the number of classes (7 in this study) to provide effective information for transiogram model estimation. The simulation accuracy results based on this sample data set are shown in Fig. 9(d). The general accuracy change trends with increasing neighborhood sizes are similar to some extent to those based on the sparse sample data set. However, for non-sectored neighborhoods, neighborhood size for the highest accuracy in simulated realizations increases a step further, to size 6, while the accuracy of optimal prediction for this neighborhood size keeps being lower than those for smaller neighborhood sizes from 2 to 5. The quadrantal neighborhood is not the best in realization accuracy or in optimal prediction accuracy if considered separately. However, considering that the neighborhood size for the highest realization accuracy and that for the highest optimal prediction accuracy are not identical, the quadrantal neighborhood is still the best choice.

Summarizing the four sets of simulation accuracy results based on the four sample data sets, we offer the following generalizations: (1) Neighborhood sizes and sectorization strongly affect simulation accuracies. (2) For non-sectored neighborhoods, with increasing neighborhood size, simulation accuracy quickly increases and then gradually deceases. However, the highest accuracy for simulated realizations and that for optimal prediction do not occur at the same neighborhood size --- the best neighborhood size for optimal prediction is always smaller than the best neighborhood size for simulated realizations. When sample density becomes sparser, the best-performed neighborhood size becomes a little larger. (3) Sectorization has a positive effect in most situations, and the quadrantal neighborhood is always the best choice. (4) Sectorization for neighborhood size 2 does not have much value due to its insufficient use of nearest data. Although octantal sectorization generally performs better than does the non-sectored neighborhood of size 8, it is also not very useful because at this neighborhood size, the simulation accuracies for realizations and optimal prediction have deteriorated greatly.

*4.2. Simulated patterns*

Simulation accuracy is just one index to reflect simulation performance. It cannot reflect whether a simulated pattern is realistic or not. Therefore, simulated patterns should be checked before making the final judgment in recommending the suitable neighborhood sizes and sector numbers for a MCRF simulation.

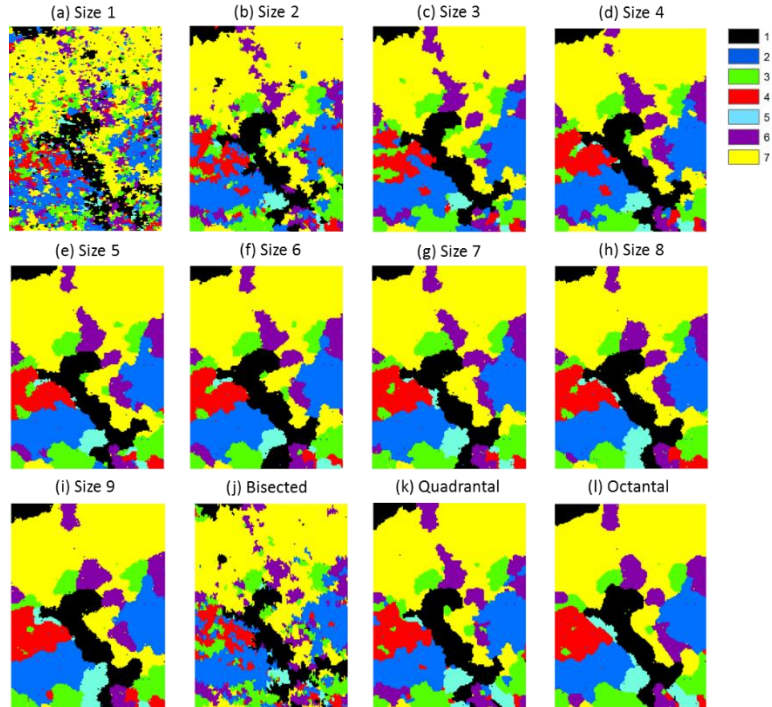

**Fig. 10.** Simulated realizations using the dense sample data set (646 data) and different non-sectored and sectored neighborhoods.

Fig. 10 shows the simulated realization maps (only the first simulated realization for each simulation case) using the dense sample data set and different non-sectored and sectored neighborhoods. The simulated patterns vary with the changes in the sizes of non-sectored neighborhoods (Fig. 10(a) to Fig. 10(i)). Compared with the original image (i.e., the reference map Fig. 7(a)), the patterns in these realization maps change with increasing neighborhood sizes from fragmentation to convergence and then to simplification (i.e., small patches are gone). The realization with the neighborhood of size 1 has a strongly fragmented pattern. The simulated pattern in the realization with neighborhood size 2 is still relatively fragmented (some minor patches occur). Correspondingly, their accuracies are also low. Thus, neighborhood sizes 1 and 2 are not suitable choices because of the pattern quality problem of the simulated realizations, despite the optimal map with neighborhood size 2 having the highest accuracy. The realizations with neighborhood sizes 3 and 4 show similar patterns as that in the original image, except for linear features that are difficult to capture using two-point statistics and limited point sample data. This may mean that neighborhood sizes larger than 2 are suitable for simulating categorical patterns. Surprisingly, a pattern-simplification trend occurs with further increasing neighborhood sizes. That is, when neighborhood size becomes large (e.g., sizes 7, 8 and 9), while the fragmentation problem is gone, the number of small patches decreases and patches become more regular, thus deviating from the normal patterns. This pattern-simplification trend corresponds to the decrease in simulation accuracy. Therefore, over-large neighborhood sizes are also not good choices due to simplification in simulated patterns. In general, for this sample data set, the neighborhood sizes 3 to 5 should be suitable choices for non-sectored neighborhoods. Checking the simulated realizations using sectored neighborhoods (Fig. 10(j) to Fig. 10(l)), one can see that the quadrantal neighborhood generates a relatively ideal realization similar to the reference map. The bisected neighborhood (also has a size of 2) does not show improvement in simulated patterns over the non-sectored neighborhood of size 2. The octantal neighborhood indicates obvious simplification (i.e., small patches are omitted) in simulated patterns. Thus, the quadrantal neighborhood should be the best choice.

For the simulated realization maps using the medium sample data set (not shown), a similar trend from fragmentation to convergence to simplification can be found in simulated patterns with increasing neighborhood sizes. Based on visual examination, we found that the sectored quadrantal neighborhood and the non-sectored neighborhoods of sizes 3 to 5 were good choices for this sample data set, and the quadrantal neighborhood was the best. This is basically the same conclusion as made based on the dense sample data set.

The simulated realization maps using the sparse sample data set and different non-sectored and sectored neighborhoods are shown in Fig. 11. Again, these simulated realization maps demonstrate similar trends, that is, with increasing neighborhood sizes the simulated patterns change from fragmentation to convergence to simplification. For the non-sectored neighborhoods, neighborhood sizes 3 to 6 are acceptable, and sizes 4 and 5 are relatively better. For sectored neighborhoods, the quadrantal neighborhood is the best, and it is also better than any non-sectored neighborhood. The visual characteristics displayed by the first simulated realizations based on this sample data set do not deviate much from those based on the dense and medium sample data sets.

The simulated realization maps using the extra-sparse sample data set and different non-sectored and sectored neighborhoods were further examined (not shown). The change trend of simulated patterns with increasing neighborhood sizes, that is, changing from fragmentation to convergence to simplification, persists. For the non-sectored neighborhoods, the patterns in the simulated realizations of neighborhood size 4 are more reasonable, and larger neighborhood sizes tend to generate simplified patterns, although neighborhood sizes 5 to 7 produce relatively higher averaged realization accuracies. For sectored neighborhoods, the quadrantal neighborhood is better than others. Due to the extra-sparseness of sample data, the simulated realizations based on this sample data set have much more uncertainty, that is, different realizations from the same neighborhood size may be quite different in patterns. Thus, for a specific neighborhood case, the single simulated realization shown here may not be representative of the patterns of other simulated realizations. Another characteristic is that the smallest class (i.e., class 5) tends to be over-represented when the neighborhood size is larger than 6. This tendency also can be seen in the simulated realizations based on denser sample data sets.

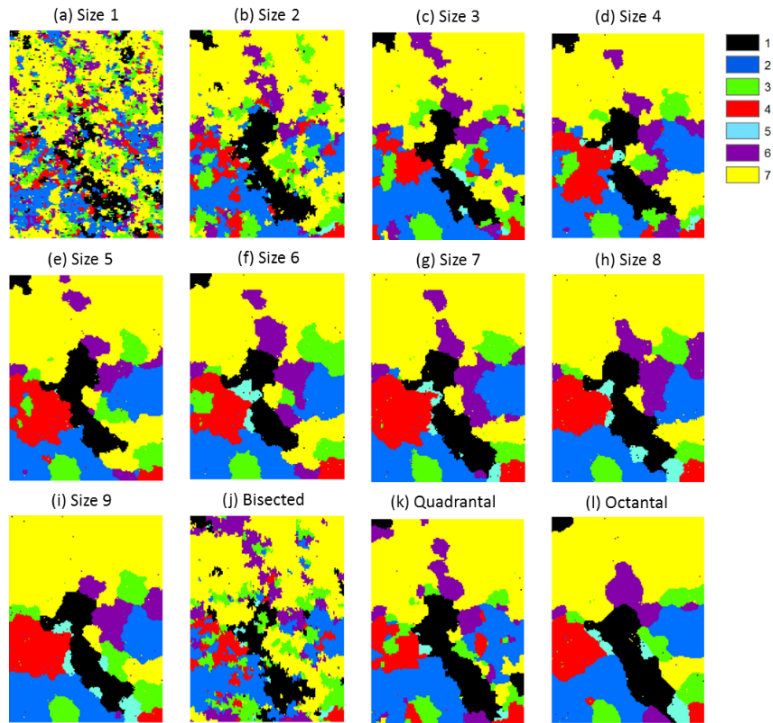

**Fig. 11.** Simulated realizations using the sparse sample data set (179 data) and different non-sectored and sectored neighborhoods.

*4.3. Effect of neighborhood deviation from conditional independence on performance*

The quadrantal neighborhood performs better than other sectored and non-sectored neighborhoods. This is reasonable theoretically. First, such a neighborhood design makes the nearest neighbors approximately meet the requirement for spatial conditional independence, based on the cardinal-neighbor conditional independence property of Pickard random fields [17, 31]. This is also the proper neighborhood size that can approximately meet the conditional independence requirement for data on a square or rectangular lattice. Second, it also reduces the data clustering (or redundancy) effect to some extent while effectively covering nearest data in the whole neighborhood search area. The bisected neighborhood has too few nearest data (only 2) and consequently cannot effectively make use of the information provided by nearest data to approach the truth of the local conditional probability distribution at an unobserved location. In addition, at upper and bottom boundaries, the neighborhood size for the bisected neighborhood may reduce to 1. These are probably the reasons why this sectored neighborhood generates fragmented patterns and shows no advantage or even worse performance compared to the non-sectored neighborhood of size 2. Although the octantal neighborhood may reduce the data clustering effect to some extent, especially when samples are dense, it apparently has larger deviation from the conditional independence of nearest neighbors. This may be the reason why it simplifies the simulated patterns (i.e., omits small patches), while it does perform relatively better than the non-sectored neighborhood of size 8 when samples are not extremely sparse.

For non-sectored neighborhoods, too small neighborhood sizes such as 1 and 2 generate fragmented patterns because they do not effectively utilize the information provided by nearest data, while large neighborhood sizes, such as sizes 7, 8 and 9, cause pattern simplification in simulated realizations because the nearest neighbors in these neighborhoods strongly deviate from conditional independence. This is why the medium neighborhood sizes (i.e., sizes 3 to 6) usually perform better. This also means that the spatial conditional independence assumption may not be proper for simplifying a complex spatial statistical model when the neighborhood size is large. In this study, the sample data are randomly distributed; and by using random paths in simulation, the simulated data are also distributed randomly. These basically ensure that the nearest data in each neighborhood are not strongly clustered at one side or in one corner, except at boundary stripes. If sample data are strongly clustered, the non-sectored neighborhoods may perform worse than shown in our analyses due to the violation of conditional independence.

It is easy to understand that low density in sample data causes low simulation accuracy. However, we find that the best neighborhood size for non-sectored neighborhoods is around 4 (i.e., 3 to 6 for averaged realization accuracy and 2 to 4 for optimal prediction accuracy), and it generally increases with decreasing sample density. This may mean that more nearest data are needed for obtaining a better probability estimate when they are located far from the unobserved location being estimated. At the early stage of the simulation of a realization, conditioning data are mostly sample data. Obviously, the accuracy of earlier simulated data decides the accuracies of later simulated data and the whole realization map.

The quadrantal neighborhood might not always be the best neighborhood. There might be some possibility that using 3, 5 or even 6 sectors might perform better than using 4 sectors (i.e., the quadrantal neighborhood) at some sample data densities, especially for generating more accurate realizations. However, it can be speculated that even if using 3 or 5 sectors sometimes may perform better than using 4 sectors, the improvement would be minimal. The reasons are apparent: The 3-sector neighborhood does not sufficiently use the information of nearest data under the condition that the conditional independence of nearest data approximately holds, while the 5-sector neighborhood has certainly deviated more from conditional independence. Sector numbers of 7 and 9 are obviously not good choices for sectoring a neighborhood for the simplified MCRF models, because neighborhoods with these two sector numbers deviate greatly from conditional independence, and consequently, similar to the octantal neighborhood (8 sectors), they would generate simplified patterns in simulated realizations.

The smallest class (i.e., class 5) is generally over-represented in simulated realizations when the neighborhood size is larger than 6 (or 7), especially when sample data are over-sparse. Transiogram models with the smallest class may not be reliable due to the insufficiency of sample data belonging to the smallest class, and large neighborhood sizes increase the chances of using unreliable transiogram models. The major reason, we speculate, may be that nearest data for these large neighborhood sizes deviate too much from conditional independence, because the simulated realizations using small neighborhood sizes do not show this feature. Further study is necessary to investigate why such a deviation can cause the over-representation of the smallest class at large neighborhood sizes.

In general, departure from conditional independence of nearest data has strong impact on simulated results of the simplified MCRF models based on the assumption of spatial conditional independence. To eliminate or minimize this impact and to make use of more nearest data in MCRF simulation, implementation of the general full solution of MCRFs or its partially simplified forms with multiple-point (or high-order) likelihoods should be promising. In addition, to completely quantify and demonstrate the impact of the neighborhood departures from conditional independence on simulated results, the best way, no doubt, is to compare the simplified MCRF models with the full MCRF models that are free of the assumption. However, the general full solution of MCRFs involves a series of multiple-point likelihoods, which are difficult to estimate from sample data and involve complex and heavy computation even if training images were available for surface landscapes. Another choice is to impose a set of adjusting parameters to the simplified MCRF models to account for the effect of neighborhood data deviation from conditional independence. However, this again may cause heavy load in computation even if there is a suitable method for imposing a set of adjusting parameters.

In real applications, incorporating auxiliary data is much more important to final simulation results than some minor improvement in simulation accuracy of the base model [22]. Although incorporating auxiliary data in MCRF cosimulation also increases computation load, it is the main way to construct integrative models for specific application purposes [40]. Therefore, looking from the angle of application practicality, the simplified MCRF solution, with a careful choice of neighborhood size and sector number, should be more useful in surface landscape modeling, especially by serving as the base model in complex integrative models. At present, we are focusing on making use of the simplified solution of MCRFs based on the spatial conditional independence assumption and the quadrantal neighborhood. Future effort may consider solving the full solution of MCRFs with multiple-point likelihoods if it is needed in some applications.

## 5. Conclusions

As a spatial Markov chain locally-conditioned through sequential Bayesian updating on nearest data within local neighborhoods, the MCRF model provides a non-linear fundamental spatial Bayesian updating solution at the neighborhood nearest data level for simulating categorical spatial variables. The general full solution of MCRFs can be visualized as a fully-related probabilistic directed acyclic graph, which is proved to be consistent with a fully-related spatial Bayesian network over the spatial nodes of nearest data within a neighborhood and the central random variable (or a neighborhood-based fully-related spatial Bayesian network). The general simplified solution of MCRFs based on the spatial conditional independence assumption of nearest data is still a neighborhood-based spatial Bayesian network, but just pairwise-related. Because a variety of specific MCRF models may be built on the general full and simplified solutions of MCRFs for various application purposes, MCRFs represent a kind of spatial Bayesian models that are neighborhood-based Bayesian networks over spatial data, although such kind of spatial Bayesian networks cannot be built through reasoning or based on intuition. The derivation method of the MCRF model provides a practical way of constructing neighborhood-based spatial Bayesian networks.

The simplified MCRF models based on the spatial conditional independence assumption of nearest data involve only spatial transition probabilities, which can be directly estimated from sample data. Thus, they can be easily implemented through sequential simulation. In order to understand the effects of different neighborhood sizes and sector numbers on simulated results using simplified MCRF models, we evaluated nine non-sectored neighborhoods (i.e., neighborhood sizes 1 to 9) and three sectored neighborhoods (i.e., 2, 4, and 8 sectors) using different densities of sample data. From the results we draw the following conclusions: (1) Neighborhood sizes and sectorization have strong impacts on simulation accuracies and simulated patterns. (2) The quadrantal neighborhood is generally the best choice, performing better than other sectored neighborhoods and non-sectored neighborhoods with regard to simulation accuracy and pattern rationality. (3) For the non-sectored neighborhoods, 3 to 5 nearest data are usually suitable choices; but when sample data are extremely sparse, relatively larger neighborhoods, such as size 6, may be a better choice for generating more accurate realizations. (4) While smaller neighborhood sizes such as sizes 1 and 2 cause simulation fragmentation, larger neighborhood sizes (usually larger than size 5) tend to simplify simulated patterns. From the accuracies and patterns of the simulated realization maps demonstrated by our testing simulation cases, we can further infer the following: (i) Departure from spatial conditional independence of nearest data should be the major reason for larger neighborhood sizes (usually larger than 5) to perform worse than medium neighborhood sizes (sizes 3 to 5) under normal sample densities. (ii) Approximately meeting the requirement for spatial conditional independence of nearest data should be the major reason for the quadrantal neighborhood to outperform the other sectored and non-sectored neighborhoods that have similar or larger sizes, while the reduction of cluster effect should also contributes. (iii) The fragmentation problem of simulated patterns using too small neighborhood sizes (i.e., sizes 1 and 2) should be caused by insufficient utilization of sample data. (iv) Given a neighborhood size, sectored neighborhoods should perform better than non-sectored neighborhoods due to the reduction of data clustering effect by sectorization, except for the neighborhood size 2 and simulations with extremely sparse sample data, whose results seem unstable.

Implementing the full solutions or partially simplified forms of MCRFs with multiple-point likelihoods may be promising. However, looking from the angles of computation efficiency and application practicality, we think that simplified MCRF models with a suitable neighborhood size and sector number should be more useful for ground surface landscape modeling, especially when incorporating auxiliary data is necessary. In addition, to avoid misunderstanding about the relationship of MCRFs with conventional Bayesian networks, we would like to clarify here that the MCRF model is a geo/spatial-statistical model and the relationship between the MCRF model and conventional Bayesian networks (including Naive Bayes) is essentially similar to the relationship between kriging and multiple linear regression.

**Acknowledgments**

This manuscript was initially prepared in 2014. Authors have the sole responsibility for all of the viewpoints presented in this paper. Partial support from U.S. NSF is appreciated.